\documentclass[aps,prc,preprint,showpacs,superscriptaddress]{revtex4}

\usepackage{graphicx}
\usepackage{afterpage}

\newcommand{\Eq}[1]   {Eq.~(\ref{#1})}

\newcommand{\agev}    {\mbox{$A$~GeV}}               %PRL notation
\newcommand{\gevc}    {\mbox{GeV$/c$}}

\newcommand{\mevc}    {\mbox{MeV$/c$}}

\newcommand{\rbt}[1]  {\mbox{\textrm{\tiny #1}}}
\newcommand{\sqrts}   {\ensuremath{\sqrt{s_{_{\rbt{NN}}}}}}

\newcommand{\der}     {\ensuremath{\textrm{d}}}

\begin{document}

%\preprint{}

%==========================================================================

\title{Antideuteron and deuteron production in mid-central Pb+Pb collisions
at 158\agev}


%\author{}
%\thanks{}
%\affiliation{}
%===========================================================================%
% List of institutions
%
\affiliation{NIKHEF,
             Amsterdam, Netherlands.}
\affiliation{Department of Physics, University of Athens, 
             Athens, Greece.}
\affiliation{E\"otv\"os Lor\'ant University,
	    Budapest, Hungary.}
\affiliation{KFKI Research Institute for Particle and Nuclear Physics,
             Budapest, Hungary.} 
\affiliation{MIT,
             Cambridge, USA.}
\affiliation{H.~Niewodnicza\'nski Institute of Nuclear Physics, Polish Academy of Sciences,
	    Cracow, Poland.}
\affiliation{Gesellschaft f\"{u}r Schwerionenforschung (GSI),
             Darmstadt, Germany.}
\affiliation{Joint Institute for Nuclear Research,
             Dubna, Russia.}
\affiliation{Fachbereich Physik der Universit\"{a}t,
             Frankfurt, Germany.}
\affiliation{CERN,
             Geneva, Switzerland.}
\affiliation{Institute of Physics, Jan Kochanowski University,
	     Kielce, Poland.}
\affiliation{Fachbereich Physik der Universit\"{a}t, 
             Marburg, Germany.}
\affiliation{Max-Planck-Institut f\"{u}r Physik,
             Munich, Germany.}
\affiliation{Inst. of Particle and Nuclear Physics, Charles Univ.,
	    Prague, Czech Republic.}
\affiliation{Nuclear Physics Laboratory, University of Washington,
             Seattle, WA, USA.}
\affiliation{Atomic Physics Department, Sofia University St.~Kliment Ohridski,
             Sofia, Bulgaria.}
\affiliation{Institute for Nuclear Research and Nuclear Energy, BAS,
             Sofia, Bulgaria.}
\affiliation{Department of Chemistry, Stony Brook Univ. (SUNYSB),
             Stony Brook, USA.}
\affiliation{National Center for Nuclear Research,
             Warsaw, Poland.}
\affiliation{Institute for Experimental Physics, University of Warsaw,
             Warsaw, Poland.}
\affiliation{Faculty of Physics, Warsaw University of Technology,
             Warsaw, Poland.}
\affiliation{Rudjer Boskovic Institute,
             Zagreb, Croatia.}
%===========================================================================%
% Author list
%
\author{T.~Anticic}
\affiliation{Rudjer Boskovic Institute,
             Zagreb, Croatia.}
\author{B.~Baatar}
\affiliation{Joint Institute for Nuclear Research,
             Dubna, Russia.}
\author{D.~Barna}
\affiliation{KFKI Research Institute for Particle and Nuclear Physics,
             Budapest, Hungary.}
\author{J.~Bartke}
\affiliation{H.~Niewodnicza\'nski Institute of Nuclear Physics, Polish Academy of Sciences,
	    Cracow, Poland.}
\author{H.~Beck}
\affiliation{Fachbereich Physik der Universit\"{a}t,
             Frankfurt, Germany.}
\author{L.~Betev}
\affiliation{CERN,
             Geneva, Switzerland.}
\author{H.~Bia{\l}\-kowska}
\affiliation{National Center for Nuclear Research,
             Warsaw, Poland.}
\author{C.~Blume}
\affiliation{Fachbereich Physik der Universit\"{a}t,
             Frankfurt, Germany.}
\author{M.~Bogusz}
\affiliation{Faculty of Physics, Warsaw University of Technology,
             Warsaw, Poland.}
\author{B.~Boimska}
\affiliation{National Center for Nuclear Research,
             Warsaw, Poland.}
\author{J.~Book}
\affiliation{Fachbereich Physik der Universit\"{a}t,
             Frankfurt, Germany.}
\author{M.~Botje}
\affiliation{NIKHEF,
             Amsterdam, Netherlands.}
\author{P.~Bun\v{c}i\'{c}}
\affiliation{CERN, 
             Geneva, Switzerland.}
\author{T.~Cetner}
\affiliation{Faculty of Physics, Warsaw University of Technology, 
             Warsaw, Poland.}
\author{P.~Christakoglou}
\affiliation{NIKHEF,
             Amsterdam, Netherlands.}
\author{P.~Chung}
\affiliation{Department of Chemistry, Stony Brook Univ. (SUNYSB),
             Stony Brook, USA.}
\author{O.~Chvala}
\affiliation{Inst. of Particle and Nuclear Physics, Charles Univ.,
	    Prague, Czech Republic.}
\author{J.G.~Cramer}
\affiliation{Nuclear Physics Laboratory, University of Washington,
             Seattle, WA, USA.}
\author{V.~Eckardt}
\affiliation{Max-Planck-Institut f\"{u}r Physik,
             Munich, Germany.}
\author{Z.~Fodor}
\affiliation{KFKI Research Institute for Particle and Nuclear Physics,
             Budapest, Hungary.}
\author{P.~Foka}
\affiliation{Gesellschaft f\"{u}r Schwerionenforschung (GSI),
             Darmstadt, Germany.}
\author{V.~Friese}
\affiliation{Gesellschaft f\"{u}r Schwerionenforschung (GSI),
             Darmstadt, Germany.}
\author{M.~Ga\'zdzicki}
\affiliation{Fachbereich Physik der Universit\"{a}t,
             Frankfurt, Germany.}
\affiliation{Institute of Physics, Jan Kochanowski University,
	     Kielce, Poland.}
\author{K.~Grebieszkow}
\affiliation{Faculty of Physics, Warsaw University of Technology, 
             Warsaw, Poland.}
\author{C.~H\"{o}hne}
\affiliation{Gesellschaft f\"{u}r Schwerionenforschung (GSI),
             Darmstadt, Germany.} 
\author{K.~Kadija}
\affiliation{Rudjer Boskovic Institute, 
             Zagreb, Croatia.}
\author{A.~Karev}
\affiliation{CERN,
             Geneva, Switzerland.}
\author{V.I.~Kolesnikov}
\affiliation{Joint Institute for Nuclear Research,
             Dubna, Russia.}
\author{M.~Kowalski}
\affiliation{H.~Niewodnicza\'nski Institute of Nuclear Physics, Polish Academy of Sciences,
	    Cracow, Poland.}
\author{D.~Kresan}
\affiliation{Gesellschaft f\"{u}r Schwerionenforschung (GSI),
             Darmstadt, Germany.}
\author{A.~Laszlo}
\affiliation{KFKI Research Institute for Particle and Nuclear Physics,
             Budapest, Hungary.} 
\author{R.~Lacey}
\affiliation{Department of Chemistry, Stony Brook Univ. (SUNYSB),
             Stony Brook, USA.}
\author{M.~van~Leeuwen}
\affiliation{NIKHEF,
             Amsterdam, Netherlands.}
\author{M.~Ma\'{c}kowiak-Paw{\l}owska}
\affiliation{Faculty of Physics, Warsaw University of Technology, 
             Warsaw, Poland.}
\author{M.~Makariev}
\affiliation{Institute for Nuclear Research and Nuclear Energy, BAS,
             Sofia, Bulgaria.}
\author{A.I.~Malakhov}
\affiliation{Joint Institute for Nuclear Research,
             Dubna, Russia.}
\author{M.~Mateev}
\affiliation{Atomic Physics Department, Sofia University St.~Kliment Ohridski,
             Sofia, Bulgaria.}
\author{G.L.~Melkumov}
\affiliation{Joint Institute for Nuclear Research,
             Dubna, Russia.}
\author{M.~Mitrovski}
\affiliation{Fachbereich Physik der Universit\"{a}t, 
             Frankfurt, Germany.}
\author{St.~Mr\'owczy\'nski}
\affiliation{Institute of Physics, Jan Kochanowski University,
	     Kielce, Poland.}
\author{V.~Nicolic}
\affiliation{Rudjer Boskovic Institute, 
             Zagreb, Croatia.}
\author{G.~P\'{a}lla}
\affiliation{KFKI Research Institute for Particle and Nuclear Physics,
             Budapest, Hungary.} 
\author{A.D.~Panagiotou}
\affiliation{Department of Physics, University of Athens,
             Athens, Greece.}
\author{W.~Peryt}
\affiliation{Faculty of Physics, Warsaw University of Technology, 
             Warsaw, Poland.}
\author{J.~Pluta}
\affiliation{Faculty of Physics, Warsaw University of Technology,
             Warsaw, Poland.}
\author{D.~Prindle}
\affiliation{Nuclear Physics Laboratory, University of Washington,
             Seattle, WA, USA.}
\author{F.~P\"{u}hlhofer}
\affiliation{Fachbereich Physik der Universit\"{a}t,
             Marburg, Germany.}
\author{R.~Renfordt}
\affiliation{Fachbereich Physik der Universit\"{a}t, 
             Frankfurt, Germany.}
\author{C.~Roland}
\affiliation{MIT,
             Cambridge, USA.}
\author{G.~Roland}
\affiliation{MIT, 
             Cambridge, USA.}
\author{M.~Rybczy\'nski}
\affiliation{Institute of Physics, Jan Kochanowski University,
	     Kielce, Poland.}
\author{A.~Rybicki}
\affiliation{H.~Niewodnicza\'nski Institute of Nuclear Physics, Polish Academy of Sciences,
	    Cracow, Poland.}
\author{A.~Sandoval}
\affiliation{Gesellschaft f\"{u}r Schwerionenforschung (GSI),
             Darmstadt, Germany.} 
\author{N.~Schmitz}
\affiliation{Max-Planck-Institut f\"{u}r Physik, 
             Munich, Germany.}
\author{T.~Schuster}
\affiliation{Fachbereich Physik der Universit\"{a}t, 
             Frankfurt, Germany.}
\author{P.~Seyboth}
\affiliation{Max-Planck-Institut f\"{u}r Physik,
             Munich, Germany.}
\author{F.~Sikl\'{e}r}
\affiliation{KFKI Research Institute for Particle and Nuclear Physics,
             Budapest, Hungary.} 
\author{E.~Skrzypczak}
\affiliation{Institute for Experimental Physics, University of Warsaw,
             Warsaw, Poland.}
\author{M.~Slodkowski}
\affiliation{Faculty of Physics, Warsaw University of Technology,
             Warsaw, Poland.}
\author{G.~Stefanek}
\affiliation{Institute of Physics, Jan Kochanowski University,
	     Kielce, Poland.}
\author{R.~Stock}
\affiliation{Fachbereich Physik der Universit\"{a}t,
             Frankfurt, Germany.}
\author{H.~Str\"{o}bele}
\affiliation{Fachbereich Physik der Universit\"{a}t,
             Frankfurt, Germany.}
\author{T.~Susa}
\affiliation{Rudjer Boskovic Institute,
             Zagreb, Croatia.}
\author{M.~Szuba}
\affiliation{Faculty of Physics, Warsaw University of Technology,
             Warsaw, Poland.}
\author{M.~Utvi\'{c}}
\affiliation{Fachbereich Physik der Universit\"{a}t,
             Frankfurt, Germany.}
\author{D.~Varga}
\affiliation{E\"otv\"os Lor\'ant University,
	    Budapest, Hungary.}
\author{M.~Vassiliou}
\affiliation{Department of Physics, University of Athens, 
             Athens, Greece.}
\author{G.I.~Veres}
\affiliation{KFKI Research Institute for Particle and Nuclear Physics,
             Budapest, Hungary.}
\author{G.~Vesztergombi}
\affiliation{KFKI Research Institute for Particle and Nuclear Physics,
             Budapest, Hungary.}
\author{D.~Vrani\'{c}}
\affiliation{Gesellschaft f\"{u}r Schwerionenforschung (GSI),
             Darmstadt, Germany.}
\author{Z.~W{\l}odarczyk}
\affiliation{Institute of Physics, Jan Kochanowski University,
	     Kielce, Poland.}
\author{A.~Wojtaszek-Szwar\'{c}}
\affiliation{Institute of Physics, Jan Kochanowski University,
	     Kielce, Poland.}

\collaboration{The NA49 Collaboration}
\noaffiliation
\date{\today \\  {\bf ..DRAFT 2.0}}
%===========================================================================

\begin{abstract}
Production of deuterons and antideuterons was studied by the NA49
experiment in the 23.5\% most central Pb+Pb collisions at the top SPS energy
of \sqrts=17.3 GeV. Invariant yields for $\overline{d}$ and $d$ were
measured as a function of centrality in the center-of-mass rapidity range $-1.2<y<-0.6$.
Results for $\overline{d}(d)$ together with previously published $\overline{p}(p)$
measurements are discussed in the context of the coalescence model. The coalescence
parameters $B_2$ were deduced as a function of transverse momentum $p_t$ and
collision centrality.
\end{abstract}

%===========================================================================% insert suggested PACS numbers in braces on next line
\pacs{25.75.Dw}
%\keywords{}
\maketitle

\section{Introduction}
Studies of antideuteron and deuteron production in heavy-ion collisions are
attractive for many reasons. Enhanced antimatter production in central nucleus-nucleus
collisions  relative to $p+p$ was proposed as one of  the experimental signatures
for formation of a new state of strongly interacting nuclear
matter - the Quark-Gluon Plasma.~\cite{antip_heinz,antip_koch,antip_ellis}.
If antibaryon abundances are not in equilibrium or not strongly affected by annihilation
after chemical freeze-out, some of the initial enhancement may survive till the final
stage of the collision process.
However, at  SPS energies baryon rich nuclear matter is created at high temperatures
($\mu_{B}\approx 250$~MeV and $T\approx$~160~MeV in central Pb+Pb collisions
at 158\agev~\cite{becat}). In a hadronic medium of such temperature and baryon density,
reaction rates involving antibaryons will be high due to annihilation and multi-pion
fusion processes~\cite{antib_bleicher,antip_rapp,antip_gavin} .  Thus any antimatter enhancement
from the partonic phase or its transition to the hadronic phase is likely to survive
only if freeze-out occurs right after the phase transition.

At the late stage of the fireball evolution,
when the hadronic matter is diluted enough such that secondary inelastic collisions cease,
the observed clusters are formed from nucleons. 
The process of nucleosynthesis in strongly interacting
systems heated to more than 100~MeV was extensively studied over the past years for
species ranging from A=2 to
A=7~\cite{clust_e878,clust_na52,clust_e802,clust_e864,clust_e877,clust_na49}.
Recently, the formation of (anti)hypernuclei was observed in Au+Au collisions by the E864
experiment at AGS energies~\cite{hyper_ags} and by the STAR collaboration
at RHIC~\cite{hyper_star}.

%At SPS energies, however, baryon-rich nuclear matter is created in the
%collisions ($\mu_{B}\approx 250$~MeV in central Pb+Pb collisions at 158\agev~\cite{becat})
%and strong antibaryon absorption is expected in such a medium.
%On the other hand, mechanisms are available for regeneration of antibaryons
%in the hadronic phase of the hot dense system either via meson-meson,
%meson-baryon interactions~\cite{antib_bleicher} or through inverse multi-pion
%annihilation~\cite{antip_rapp}. Therefore antibaryon production is a valuable probe
%of the reaction dynamics and space-time evolution of the created
%fireball~\cite{antip_gavin}.
%At the late stage of the fireball evolution, when the source is rather diluted,
%the (anti)baryon zoo copiously produced
%in A+A interactions is enriched by (anti)nucleon clusters. The mechanism of nuclear
%synthesis in strongly interacting systems heated to more than 100 MeV
%was extensively studied over the past years for species ranging from $A$=1 to
%$A$=7~\cite{clust_e878,clust_na52,clust_e802,clust_e864,clust_e877,clust_na49}.
%The formation of even more exotic objects, i.e., (anti)hypernuclei,
%was observed in A+A collision by the E864 experiment at AGS energies~\cite{hyper_ags}
%and by the STAR collaboration at RHIC~\cite{hyper_star}.

In a simplified coalescence approach~\cite{coal1,coal2}, (anti)nucleon bound
states are formed from (anti)nucleons which are close in momentum and configuration space. Their production is usually characterized by the coalescence factor $B_A$
which relates the invariant yield of a cluster of size $A$ to the $A^{th}$ power
of the proton yield.
The $B_2$ parameter for deuterons was found to be approximately constant
in heavy-ion collisions at beam energies below 1\agev\ as well as in all proton-induced reactions. 
However, at higher collision energies (more than several GeV per nucleon)
the diameter of the source created in central collisions is much larger than the deuteron size due to expansion of the reaction zone, and the cluster
production process becomes sensitive to the details of the phase-space distribution
of the constituents.  
In particular, a strong collective radial flow (the average transverse velocity
is about 0.5$c$ at SPS energies~\cite{na49_hbt}) produces correlations between the 
production point in space-time and the momentum of the clusters. Thus, an interplay
between the flow velocity profile and the nucleon density distribution may result in
different shapes of transverse momentum spectra for composites and single
nucleons~\cite{slope_polleri}. A comparative analysis of the production
of clusters made of nucleons and antinucleons might therefore provide insight
into the details of the fireball evolution and the nucleon
phase-space distribution at kinetic freeze-out. 
The production rate of composites should depend on the spatial
distribution of their constituents. If the $\overline{p}$ suffer from annihilation
in the dense baryonic matter, this could therefore be reflected in the
impact parameter dependence of the $\overline{d}$ yield~\cite{antib_bleicher_2}.
Moreover, because of the additional annihilation cross section, $\overline{d}$ are
expected to decouple from their source at much lower hadron density than $d$
and therefore their freeze-out volume is expected to be larger.
In summary, the study of $\overline{d}$ and $d$ production provides an
alternate method for extracting information about the source
size~\cite{mrozhin,coal_heinz} which is complementary to the femtoscopy approach.  

Recently, NA49 reported~\cite{na49_ppbar} on $\overline{p}$
production at midrapidity in centrality selected Pb+Pb collisions at
158\agev.
In the following we extended this study of antimatter to the
investigation of heavier systems.
This paper reports our measurement of $\overline{d}$ and $d$ invariant
yields around midrapidity in the same reaction.

The paper is organized as follows. The next section gives an overview
of the NA49 experimental set-up. Details of the data analysis procedure
are described in Section~III. In Section~IV we present and discuss the results on
$\overline{d}$ and $d$ production in mid-central Pb+Pb collisions.
A summary is given in the last section.

\section{Experimental setup}

The NA49 apparatus was designed as a large acceptance
magnetic spectrometer to study nucleus-nucleus collisions in the SPS
energy range. The NA49 setup covers about 50\% of the final
state phase space for Pb+Pb reactions at a beam energy of 158\agev.
The detector provides precise tracking and robust particle
identification. A full description of the apparatus
components can be found in Ref.~\cite{na49_setup}.

The primary lead beam was extracted from the CERN Super Proton Synchroton (SPS)
and delivered to NA49 through the H2 beam line in the SPS North Area.
The intensity of the beam (delivered in spills 4.8~s long) was
typically $10^5$/s.
The beam was defined within the experiment by
a set of Cherenkov and scintillation counters placed along 35~m
of the beam path upstream from the nominal target position.
The transverse diameter of the beam spot on the target,
as measured by three multi-wire proportional beam position
detectors (BPDs), was about 1~mm.
The target was a lead foil of 337~mg/cm$^2$ thickness
(1.5\% of a Pb interaction length).
The interaction trigger required a valid beam signal plus a collision
centrality tag based on the measurement of projectile spectator energy
registered in the zero-degree calorimeter VCAL which was placed 23~m
downstream of the target.

Charged reaction products originating from the interaction vertex
are recordeed by four gas Time Projection Chambers (TPCs).
Two Vertex TPCs (VTPC), serving mainly for momentum analysis,
are located inside two superconducting dipole magnets,
which provide a magnetic field with a total bending power of
about 9~Tm.
The two Main TPCs (MTPC), with about 27~m$^3$ gas volume each,
are positioned behind the magnets on each side of the beam line.
These TPCs are optimized for energy loss ($dE/dx$) measurements
with a precision of 4\% for minimum ionizing particles ({\it mip}).

Two Time-of-Flight (TOF) arrays (walls), positioned behind the MTPCs
at $\approx$~14~m from the interaction vertex,
are the main detector systems used in this analysis for the identification
of charged hadrons of momenta up to 10~\gevc. The TOF setup provides
rapidity coverage of about 0.8 units in the mid-rapidity
region and transverse momentum coverage up to $p_t$~=~2.5~\gevc.
Each wall consists of 891 pixels made of 2.5~cm thick plastic
scintillators with transverse dimensions of 6, 7 and 8~cm~(horizontal)
by 3.4~cm~(vertical). Each pixel is viewed by one phototube of 1.4 inch
diameter glued to the side of the pixel. The phototube outputs (typically
1.5~Volts high for a charge~=~1 {\it mip} passing through the pixel) are
divided into two equal parts and digitized by TDCs and ADCs for time-of-flight
and energy loss measurements, respectively.
The least count of the FASTBUS LeCroy TDCs was set to 25~ps.
Pixels are assembled in groups of 11 (as stacks in the vertical direction)
in light-protected cassettes.

A 0.5 mm thick quartz Cherenkov counter $S_{1}$ placed 34~m upstream
of the target provided the common start signals and the gates to the
TDCs and ADCs, respectively. For a beam of lead nuclei $S_1$ has an intrinsic
time resolution about 30~ps.
The overall time resolution achieved with the TOF system was
$\sigma\approx 60$~ps and allows for a 5$\sigma$ $p/d$
separation up to a momentum of $p$~=~10~\gevc.

\section{Data analysis}

\subsection{Data sets}
A high statistics sample of Pb+Pb collisions at beam energy of 158\agev\ was collected
during the year 2000 run period.
A total of $2.4\cdot 10^6$ events were recorded with an online
event trigger selecting the 23.5\% most central collisions.

\begin{table}[tbh]%[H] add [H] placement to break table across pages
\caption{\label{table:tab1} Summary of the data sets used in the analysis.
The number of events employed are given together with the fraction of the total
cross-section (in percent) and corresponding average number of wounded nucleons
$\langle N_w \rangle$ per event derived from the VENUS model.}
\begin{ruledtabular}
\begin{tabular}{lll}
        \multicolumn{1}{c}{centrality}
        &\multicolumn{1}{c}{$\langle N_w \rangle$}
        &\multicolumn{1}{c}{$N_{events}$} \\
\hline
0-23.5\%& 265 &2,400,000\\
0-12.5\% & 315 & 1,240,000\\
12.5-23.5\% & 211 & 1,160,000\\
\end{tabular}
\end{ruledtabular}
\end{table}
To study the centrality dependence of the $\overline{d}$ and $d$ yields the
data set was divided into two centrality classes
(see Table~\ref{table:tab1} for detail). For each centrality selection,
the corresponding number of wounded nucleons $N_w$ as well as the percentage
of the total cross-section were obtained from the Glauber model approach using
a simulation with the VENUS event generator~\cite{venus,na49_centrality}.

\subsection{Time of flight reconstruction}
The analysis procedure starts with the calibration of all TOF pixels:
time-zero offsets $T_0$, ADC gains, and pedestals were determined.
All reconstructed tracks in the TPCs
with assigned momentum were extrapolated towards the TOF detectors
and all possible associations  between the track extrapolations
and the calibrated TOF hits were found.
For each TOF hit the raw TDC count was then corrected for the light propagation
time in the scintillator according to the position of the point from which
the light originated, and for the amplitude-dependent time-walk effect in the discriminator.
The correction was as large as 50~ps/cm and 30~ps for the light
propagation time and time-walk effect, respectively.
 Finally, an event-by-event global time offset
($\approx$40-50~ps on average) was determined as the mean
of the distribution of the differences between the measured and calculated arrival time
for pions selected by their energy loss $dE/dx$ in the TPCs 
(15 to 30  $\pi$'s per wall in an event depending on the collision centrality).
The achieved overall time resolution of the TOF system was about 60~ps.

The squared mass $m^2$ was calculated from the reconstructed  momentum $p$,
the flight path to the TOF detector $l$ and the measured time-of-flight $t$ as
\begin{equation}
  \label{eq:mass2}
  m^2=\frac{p^2}{c^2} \left( \frac{c^2t^2}{l^2}-1 \right)
\end{equation}
where $c$ denotes the speed of light.

\subsection{Event and track selection}

In the following we discuss the event selection criteria as well as the track
and TOF hit quality cuts which were applied during the analysis.
In order to eliminate the background from non-target interactions
an offline cut on the vertex $z$ coordinate along the beam line was imposed.
The difference between the $z$-position reconstructed from tracks
and the nominal (survey) target position was required to be
less than 1~cm ($\pm 4\sigma$ deviation, with $\sigma$ being the RMS of the
vertex $z$ distribution). The fraction of events rejected by this cut
was about 0.5\%.

Track candidates were required
to have a valid momentum fit result and $dE/dx$ measurement.
In order to reduce the contamination by background tracks
from secondary interactions in the material or decays
inside the TPC, the following track quality criteria were applied:
\begin{enumerate}
\item
Tracks were accepted if they had points in the MTPC and in at least one of 
the VTPCs, and the length of the MTPC track segment was longer than 1.5~m.
\item
Both vertical and horizontal
deviations from the vertex position,
after back extrapolation to the $z$ position of the target,
were required to be less than 1~cm.
\end{enumerate}
An additional set of quality cuts was imposed to reject those TOF hits
that have poorly reconstructed time of flight:
\begin{enumerate}
\item
Candidates were rejected if two or more tracks in the event hit the same
scintillator.
\item
A distance greater than 1~mm from the pixel's edges was
required for the position of a TOF hit to account for mismatches
between TPC tracks and TOF signals due to multiple scattering and TPC-TOF
misalignment.
\item
A cut on the energy deposited in a scintillator was applied
to reduce background under the (anti)deuteron mass peak due to false TOF hits
and interactions of $\gamma$s associated to the interactions. The ADC value normalized to minimum
ionizing particles was required to be within $0.8 < ADC/ADC_{mip} < 1.4$.
\end{enumerate}
The fraction of the TOF hits remaining after all the quality cuts
was found to be 67\%.

\subsection{Identification of deuterons and antideuterons}

Particle identification is done in two steps by combining measurements
of time-of-flight from the TOF detector and $dE/dx$ from the MTPCs.
\begin{figure}
\includegraphics[width=0.7\linewidth]{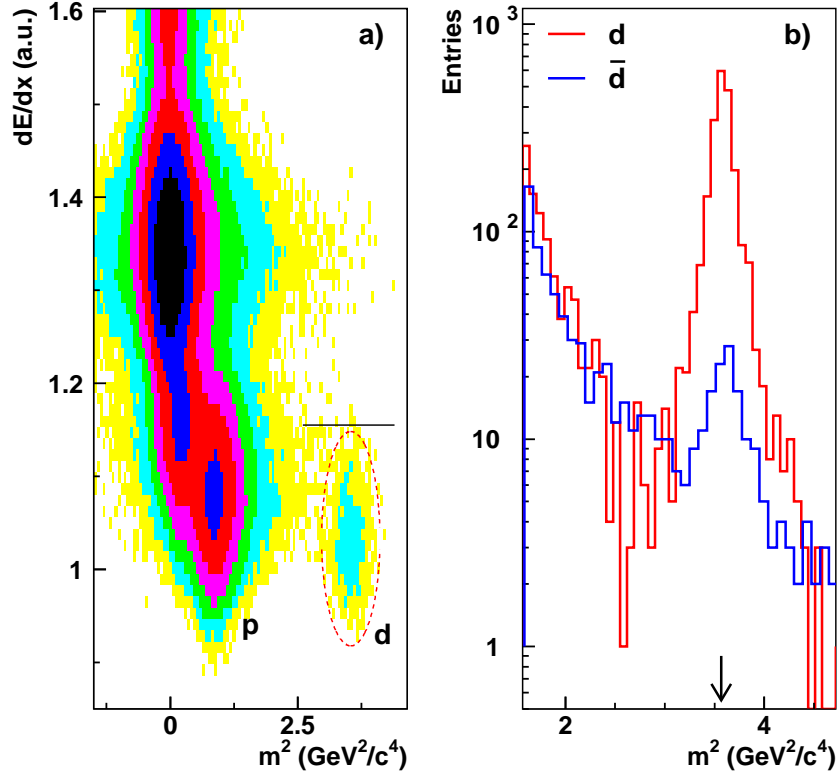}
\caption{(Color online) a) $dE/dx$ versus $m^2$
for positively charged particles in the momentum interval $7<p<8$~\gevc. The dashed ellipse
illustrates the region populated by deuterons; 
the PID upper-limit $dE/dx$ cut is indicated by the horizontal line.
b) $m^2$-distributions for both charges around the $m^2_d$
position for momenta from 4 to 10~\gevc\ ($dE/dx$ PID cut for
$d (\overline{d})$ was applied). The arrow indicates the nominal $m^2$ position
for $d(\overline{d})$.}
\label{fig:pid_deut}
\end{figure}
First, the allowed values of $dE/dx$ were limited to
$$
dE/dx<dE/dx_{BB}(p)+k\sigma(p),
$$
 where $dE/dx_{BB}(p)$ is the most probable value as taken from
a momentum dependent parameterization of the Bethe-Bloch curve for deuterons,
and $\sigma(p)$ is the corresponding resolution in $dE/dx$. The parameter $k$
%was chosen to be equal
was set to 3 and 2 for $d$ and $\overline{d}$, respectively. As illustrated
in Fig.~\ref{fig:pid_deut} (left panel) for the momentum bin $7<p<8$~\gevc\ such a cut reduces
the contribution of pions and kaons to the $\overline{d}$~($d$).
The resulting $m^2$-distributions for positively and negatively charged particles
after applying the previously described quality and $dE/dx$ PID cuts show
clear $d$ and $\overline{d}$ peaks (Fig.~\ref{fig:pid_deut}, right panel).
The entire track sample of $\overline{d}$($d$) was divided
into transverse momentum bins of 0.15~\gevc\ width; the three highest $p_t$ bins
were combined into a wider one because of lack of statistics.
In each $p_t$~bin candidates are selected by a $\pm 3\sigma$ window in $m^2$ around the nominal
position. The raw yields were determined by summing the histogram
bin contents within the mass window. The background contamination
under the $\overline{d}$~($d$) peak was evaluated by fitting the $m^2$-distribution
to a sum of a Gaussian signal and an exponential background.
\begin{figure}
\includegraphics[width=0.7\linewidth]{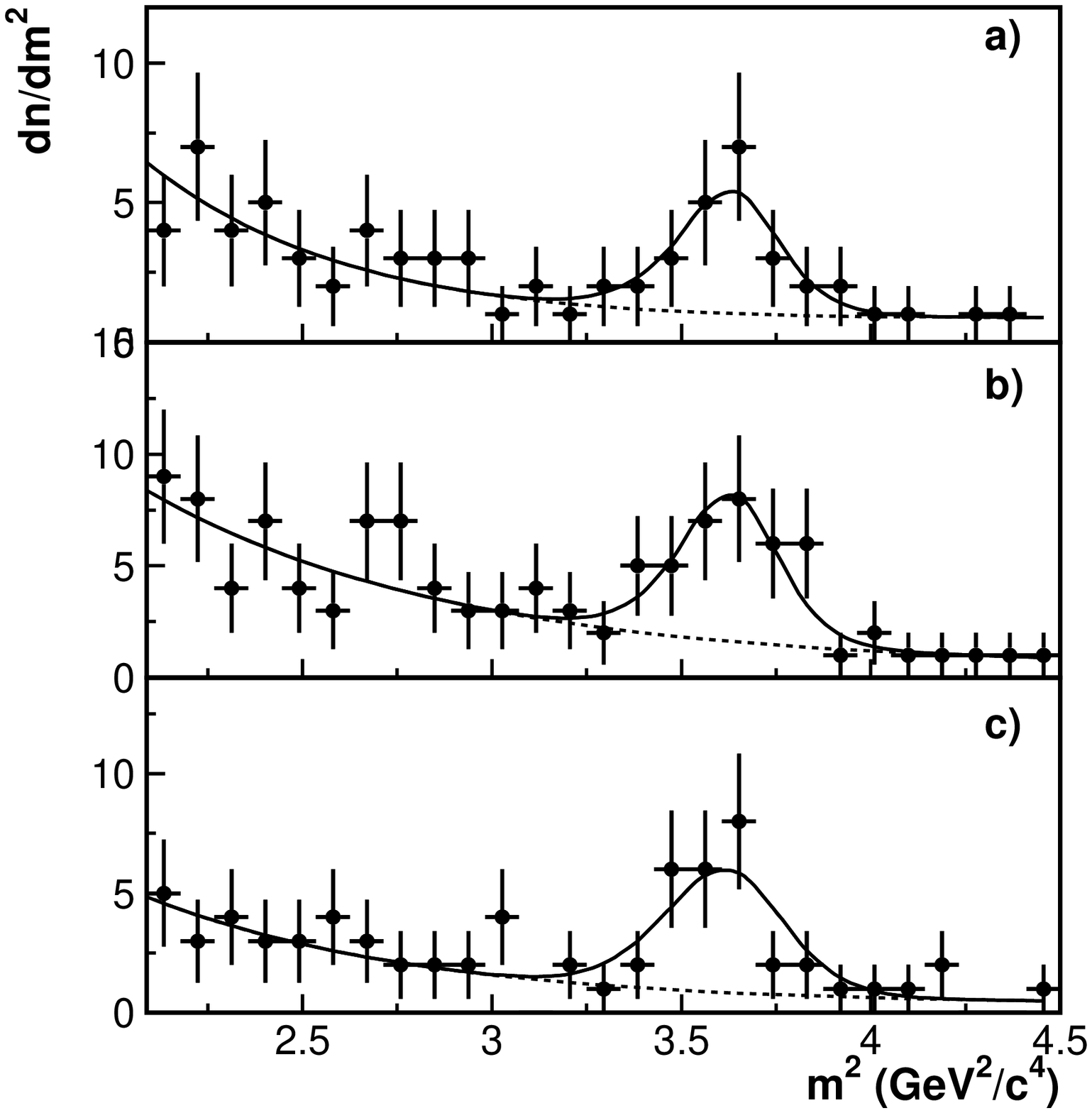}
\caption{Mass-squared distributions of $\overline{d}$ for rapidity  $-1.2<y<-0.6$ 
in several transverse momentum bins: a) $0<p_t<0.3$~\gevc, b) $0.3<p_t<0.45$~\gevc,
c) $0.45<p_t<0.6$~\gevc. Solid lines represent fits of a sum 
of a Gaussian signal and an exponential background.}
\label{fig:m2_dbar}
\end{figure}
Figure~\ref{fig:m2_dbar} shows mass squared distributions
along with the combined (Gauss + exponential) fits for $\overline{d}$
for three $p_t$ bins.
The background contamination is weakly dependent on
$p_t$ and amounts to less than 3\% for deuterons, but reaches
$\approx$~45\% for $\overline{d}$.
The statistical error of the raw $\overline{d}$ yields was calculated as
$\sqrt{N}=\sqrt{N_{\overline{d}}+N_{bkg}}$, where $N_{\overline{d}}$ and
$N_{bkg}$ are the number of $\overline{d}$ and background counts, respectively.

\subsection{Corrections}
The raw yields were weighted by correction factors which account
for the geometrical acceptance and for the losses due to the applied
track quality and $dE/dx$ PID cut.

\begin{figure}
\includegraphics[width=0.6\linewidth]{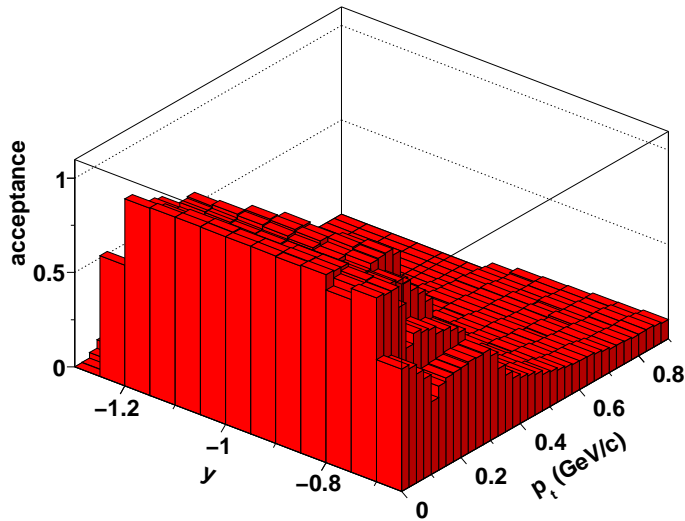}
\caption{(Color online) Geometrical TOF acceptance as a function of rapidity
$y$ and transverse momentum $p_t$ for $d$ at 158\agev.}
\label{fig:dbar_accept}
\end{figure}

Correction factors for the limited geometrical coverage of the detector system
were obtained from a GEANT-based Monte Carlo simulation. 
Figure~\ref{fig:dbar_accept} shows the NA49 TOF acceptance for $d$($\overline{d}$)
in terms of transverse momentum $p_t$ and center-of-mass rapidity $y$. The bin size was 
taken to be $\Delta y$=0.1 by $\Delta p_t$=25~\mevc, and all the bins
at the edges of the acceptance were rejected to avoid a very large correction factor.
The acceptance drops down to about 6\% at $p_t> 0.6$~\gevc.
Monte Carlo studies with simulated tracks embedded into
real data showed that the tracking reconstruction
efficiency in the phase-space region covered by the TOF detector is above 98\%.
All remaining corrections were determined from the data in bins of $p_t$.
For each centrality bin the corrections for multihits were obtained by determining 
the fraction of tracks rejected by the requirement that only a single one hits
a scintillator. This correction shows less than 3\% variation with centrality
and is on average 11\% and 8\% for the 0-12.5\% and 12.5-23.5\% most central events,
respectively. The losses due to the geometrical fiducial cut
(1~mm edge cut) were determined averaged over all centralities.
This correction accounts for about 9\% losses.
The corrections (about 15\%) due to the pulse height constraint ($0.8 < ADC/ADC_{mip} < 1.4$)
were determined for the deuteron sample and applied for both $\overline{d}$ and $d$.
The corrections for the losses due to the DCA cut
($\approx 2\%$) were also determined from the DCA 
distributions of the deuteron candidates.

\subsection{Systematic errors}

The systematic errors of the $d$ and $\overline{d}$
yields originate mainly from the uncertainties of particle identification
and efficiency correction factors.
In general, each particular contribution to the overall systematic error
was estimated by varying the characteristic selection
criteria, i.e., by loosening some cut or making it tighter.
The contributions of the following selection criteria were studied:
the $dE/dx$ identification cut, the pulse height and geometrical cuts for
selecting TOF hits and the DCA cut for selecting tracks.
The uncertainty arising from the PID procedure for $\overline{d}$
(including background subtraction) introduces a systematic error of the order of 13\%.
The DCA cut contributed $\approx$~5\% to the overall systematic error.
The error associated with the TOF efficiency correction for $\overline{d}$
was found to be 9\%.

The overall systematic error, obtained as the quadratic sum
of all contributions, was estimated to be 17\% for $\overline{d}$ and
10\% for $d$.

\section{Results and discussion}

\subsection{Transverse momentum distributions}

\begin{figure}
\includegraphics[width=0.7\linewidth]{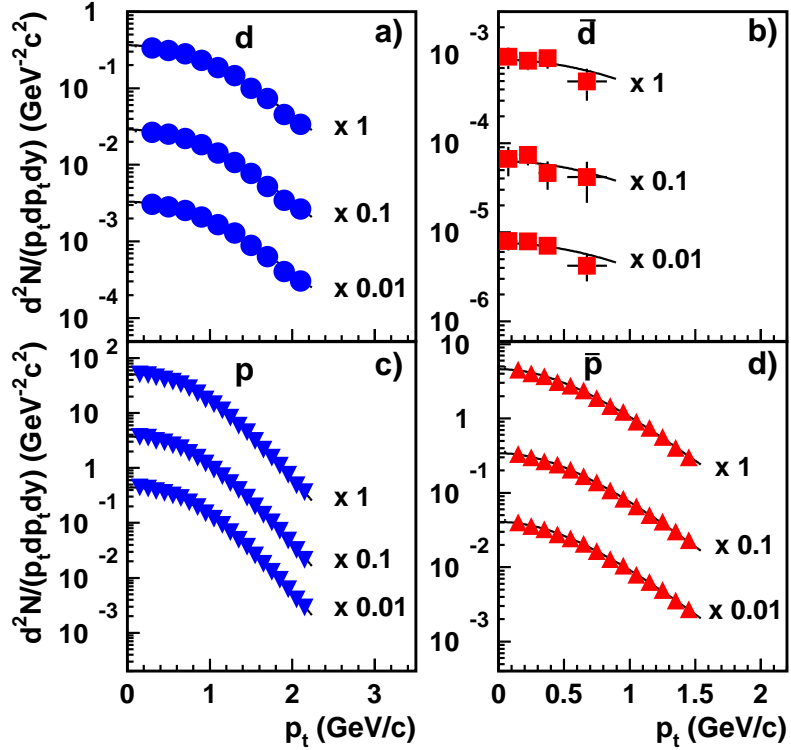}
\caption{(Color online) The transverse momentum spectra of deuterons (a), antideuterons (b), 
protons (c) and antiprotons (d) in centrality
selected Pb+Pb collisions at 158\agev. Only statistical errors are shown.
The curves show exponential fits to the data using \Eq{eq:b1}. Results for the 0-12.5\%
most central collisions are shown to scale, results for the centrality intervals
12.5~-~23.5~\% and 0~-~23.5~\% are divided by factors $10$ and $10^2$ respectively.}
\label{ddbar_spectra}
\end{figure}

Figure~\ref{ddbar_spectra}~(a,b) shows the invariant yields of $d$ and
$\overline{d}$ for centrality selected Pb+Pb collisions
as a function of transverse momentum $p_t$. The yields are averaged
over the center-of-mass rapidity interval $-1.2 < y < -0.6$.
The data for the most central bin are shown to scale,
while each successive distribution was divided by an additional factor of ten for clarity
of presentation.
%%%%
%%%% Vadim. added 3.08.2010
We have also analyzed $p$ and $\overline{p}$ yields in the same centrality selected event
samples. A detailed description of the analysis procedure for (anti)protons can be found
elsewhere~\cite{na49_ppbar}. The invariant distributions for $p$ and $\overline{p}$
measured in the rapidity interval $-0.5<y<-0.1$ are shown in Fig.\ref{ddbar_spectra}~(c,d).
The yields agree with the previously published data~\cite{na49_ppbar} within 12\%,
the observed difference is mainly due to feed-down corrections from weak decays which
are now based on the recent experimental data on $\Lambda$($\overline{\Lambda}$) production
from the NA49 experiment~\cite{na49_lam_energy,na49_lam_size}. Numerical values of
the transverse momentum spectra are given in Table~\ref{table:tab2}.  
%%%%

Calculations using the microscopic transport model UrQMD~\cite{urqmd}
suggest that, due to annihilation in dense baryon matter,
a considerable difference between the $p_t$ distributions of
anticlusters compared to those of clusters is expected.
If annihilation of antimatter in the fireball medium affects
$\overline{d}$ production, then one would expect $\overline{d}$
losses to be larger at midrapidity and at low $p_t$~\cite{anti_loss},
resulting in a hardening of the $p_t$ spectra of $\overline{d}$
compared to those for $d$. On the other hand, the NA49 measurements
of mid-rapidity $\overline{p}$($p$) spectra~\cite{na49_ppbar} show little 
difference between transverse distributions of $p$ and $\overline{p}$
for all centralities, leading to the conclusion that the $p_t$ dependence 
(if any) of annihilation losses for antiprotons is small at 158\agev.
A similar behavior is also observed for the lambdas and antilambdas~\cite{na49_lam_size}.

The experimental deuteron distributions were fitted to an exponential function in $m_t$
(shown by curves in Fig.~\ref{ddbar_spectra}~(a))

\begin{equation}
\label{eq:b1}
\frac{1}{p_t}\frac{d^2N}{dp_tdy}=
\frac{dN/dy}{T(m+T)}\exp{\left(-\frac{m_t-m}{T}\right)}
\end{equation}
where $dN/dy$ and $T$ are two fit parameters, $m_t = \sqrt{p_t^2 + m^2}$ is the
transverse mass and $m$ the deuteron rest mass.

The small available statistics and resulting limited $p_t$ range for $\overline{d}$
data makes the determination of the slope parameters for $\overline{d}$
less accurate. Thus only the $dN/dy$ parameter was allowed to vary
when the $\overline{d}$ spectra were fitted and merely a qualitative
comparison of the shape of the spectra is possible.
As one can see in Fig.~\ref{ddbar_spectra}~(b), the fits of
$\overline{d}$ spectra with slope parameters fixed to the values obtained from
the $d$ distributions from the same collision centrality
(Fig.~\ref{ddbar_spectra}~(a)) appear to describe
the $\overline{d}$ data well.
\begin{figure}
\includegraphics[width=0.6\linewidth]{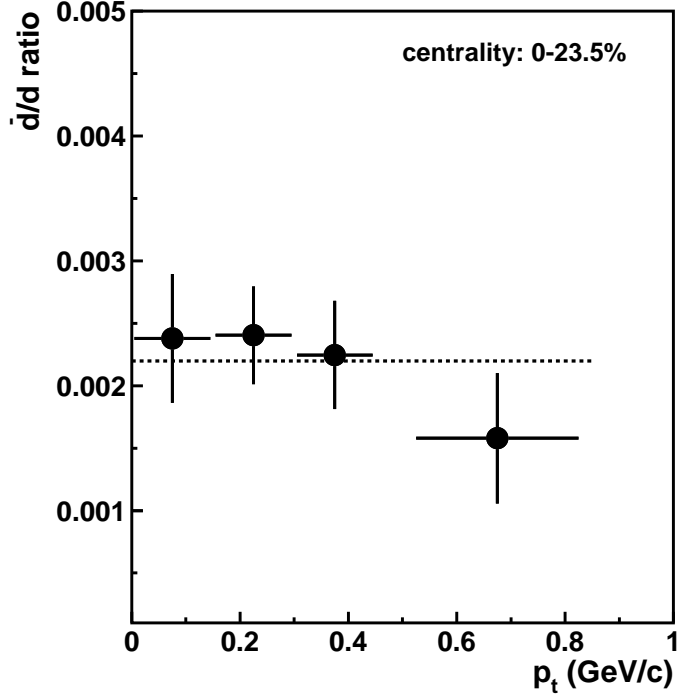}
\caption{ $\overline{d}/d$-ratio as a function
of $p_t$ in 0-23.5\% central Pb+Pb collisions at 158\agev. The dashed line shows
a fit to a constant.}
\label{fig:dbar_d_ratio}
\end{figure}
In Fig.~\ref{fig:dbar_d_ratio}, the $\overline{d}/d$ ratio for the 0-23\% centrality
bin is plotted as a function of $p_t$. The ratio shows little variation with $p_t$
within the errors. A fit with a constant yields an overall $\overline{d}/d$ ratio
of $(2.2 \pm 0.2)\cdot 10^{-3}$.  
Since no difference in shapes of $\overline{d}$ and $d$ spectra
was observed, a likely conclusion is that the dynamical properties of the formation
process are the same for $d$ and $\overline{d}$ production. There is also no indication
of extra reduction of the $\overline{d}$ yield at low $p_t$ due to annihilation.

The fit parameters $dN/dy$ for the $\overline{d}$ and
$d$ spectra in different centrality bins are tabulated
in Table~\ref{table:tab2}.
The fraction of the yield integrated using Eq.~\ref{eq:b1} over the unmeasured
$p_t$ range is $\approx$7\% for $d$ and 55\% for $\overline{d}$.

The slope parameter $T$ for deuterons is $406\pm 12$~MeV
and $391\pm 12$~MeV for the 0~-~12.5~\% and 12.5~-~23.5~\% most central
collisions, respectively. It tends to increase with increasing centrality
indicating increase of collective transverse flow.   
In Ref.~\cite{clust_na49} we reported yields and slopes for
deuterons from the 1996 minimum bias data set of Pb+Pb collisions
at 158\agev, subdivided into 6 centrality bins. These measurements
for deuterons were obtained in the rapidity range $-0.9 < y < -0.4$,
and only tracks with $p_{x} > 0$ were used.
Taking into account the different rapidity intervals
for both measurements (an increase of 15\% in $dN/dy$ and a decrease of about
5\% in $T$ is expected when going from $y$=~$-0.65$ to $y$=~$-0.9$),
the overall agreement of the present results with our previously
published data is satisfactory.

\begin{table}[tbh]%[H] add [H] placement to break table across pages
\caption{\label{table:tab2} $dN/dy$ for $d$ and $\overline{d}$ obtained
from fits with \Eq{eq:b1} in centrality selected Pb+Pb collisions
at 158\agev\ in the rapidity interval $-1.2<y<-0.6$. Errors are statistical only.}
\begin{ruledtabular}
\begin{tabular}{lll}
        \multicolumn{1}{c}{centrality}
        &\multicolumn{1}{c}{deuterons}
        &\multicolumn{1}{c}{antideuterons} \\
\hline
0~-~12.5~\%& $0.33\pm 0.02$ &$(8.1\pm 1.1)\cdot 10^{-4}$\\
12.5~-~23.5~\% & $0.25\pm 0.02$ & $(5.6\pm 1.0)\cdot 10^{-4}$\\
0~-~23.5~\% & $0.3\pm 0.01$ & $(6.9\pm 1.0)\cdot 10^{-4}$\\
\end{tabular}
\end{ruledtabular}
\end{table}
% slopes 406 +/- 2 (0-12%), 391 +/- 2 (12-23%)
%    400 +/- 1 (0-23%)
The invariant $p_t$ spectra of $\overline{d}$($d$) are harder than those of 
$\overline{p}$($p$)~\cite{na49_ppbar} (typical $T_p\approx290-300$ MeV).
As was pointed out in~\cite{slope_polleri,coal_heinz}, the characteristic
change of the slope with the particle mass is sensitive to the interplay
between the density and flow velocity profiles in the source at freeze-out.
The observed increase of the inverse slope parameter $T$ of about 30\% for $d$ 
supports a box-like density profile.

\subsection{Rapidity distributions and total yields}

\begin{figure}
\includegraphics[width=0.7\linewidth]{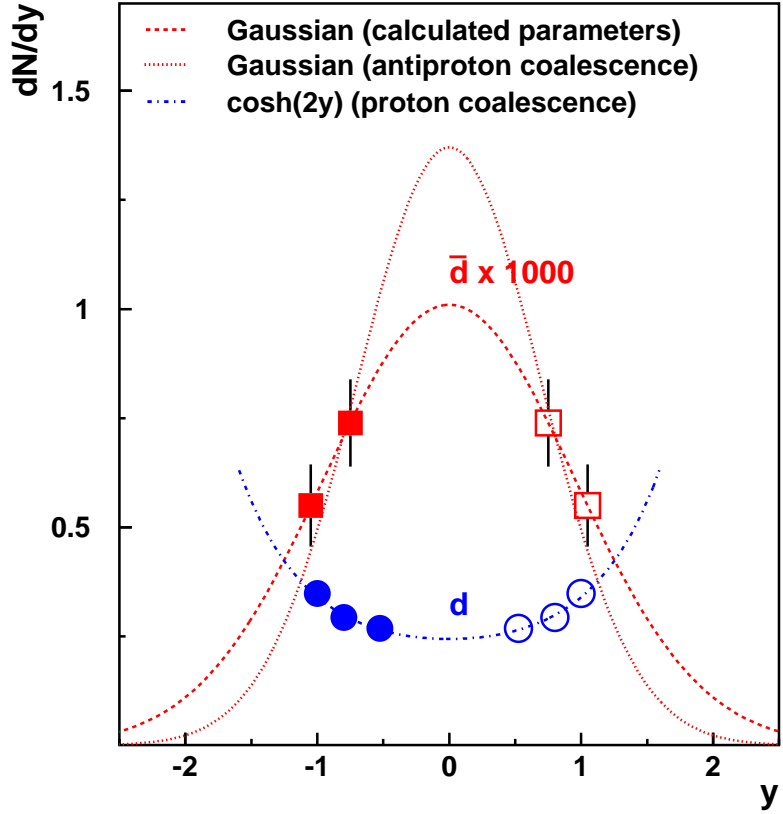}
\caption{(Color online) Rapidity distributions
for $\overline{d}$ (squares) and $d$ (circles) produced in the
23.5~\% most central Pb+Pb collisions at 158\agev.
The solid symbols indicate the measured data points, the open symbols
are reflected at midrapidity. The curves represent the results of different
fits motivated by a coalescence approach (see text).}
\label{fig_rap}
\end{figure}

Yields of $d$ and $\overline{d}$ as a function of rapidity
for the 23.5~\% most central Pb+Pb collisions
are shown in Fig.~\ref{fig_rap}.
There is a difference between the shapes of the rapidity
spectra for $d$ and $\overline{d}$. The observed distinction
can be traced back to the longitudinal distributions of their constituents.
Data on $p$ and $\overline{p}$ production in centrality selected 
Pb+Pb collisions at 158$A$~GeV were recently published by NA49~\cite{na49_ppbar_dedx}.
For collisions of all centralities, the proton rapidity distributions have a dip at
midrapidity in contrast to the peak observed for antiprotons.
Since the abundance of (anti)clusters (of mass number $A$) is
proportional to the (anti)nucleon phase-space density raised to
$A^{th}$ power, the effect of the difference in the shapes of the rapidity 
distributions is even more pronounced for the composites.
In Fig.~\ref{fig_rap} two Gaussian distributions are plotted for $\overline{d}$.
The dotted line represents the simplest expectation from the coalescence model,
namely a Gaussian with a width which
is smaller by a factor of $\sqrt{2}$ than that for antiprotons for the centrality
class 0-23\%~\cite{na49_ppbar_dedx}. The integral of this distribution
is equal to that of the second Gaussian (shown by the dashed line), whose parameters,
i.e. the width and total yield were calculated based on the measured
points (two unknown parameters and only two measured points). The calculated parameters 
were found to be $2.4\cdot 10^{-3}$ for the $4\pi$-yield and $0.95$ for the width.
The width of the rapidity distribution for antideuterons is close
to that observed for antiprotons ($\sigma_{\overline{p}}\approx0.93-1.03$)
in the centrality range under study~\cite{na49_ppbar_dedx}. For $d$ production
the coalescence model predicts a parabolic dependence on rapidity based on
that measured for protons \cite{na49_ppbar_dedx}.
The integral of the parabolic parameterization for $d$ yields
$2.8\pm0.1$.

Rough estimates for the total (anti)cluster yields
were made in the framework of a statistical hadron gas model (HGM)~\cite{becatini_private}.
A total yield of 2.5 and $4.6 \cdot 10^{-3}$ was predicted
in the 5\% most central Pb+Pb collisions at 158\agev\ for $d$ and $\overline{d}$,
respectively. Assuming the yield of clusters to scale with
the number of wounded nucleons ($N_w$ changes by a factor of about 1.4
from 5\% to 23.5\% centrality of collisions), the difference between data and the HGM
estimates is below 40\% for both $d$ and $\overline{d}$. 
 
\subsection{Centrality dependence of $\overline{d}$ and $d$ yields}
\begin{figure}
\includegraphics[width=0.7\linewidth]{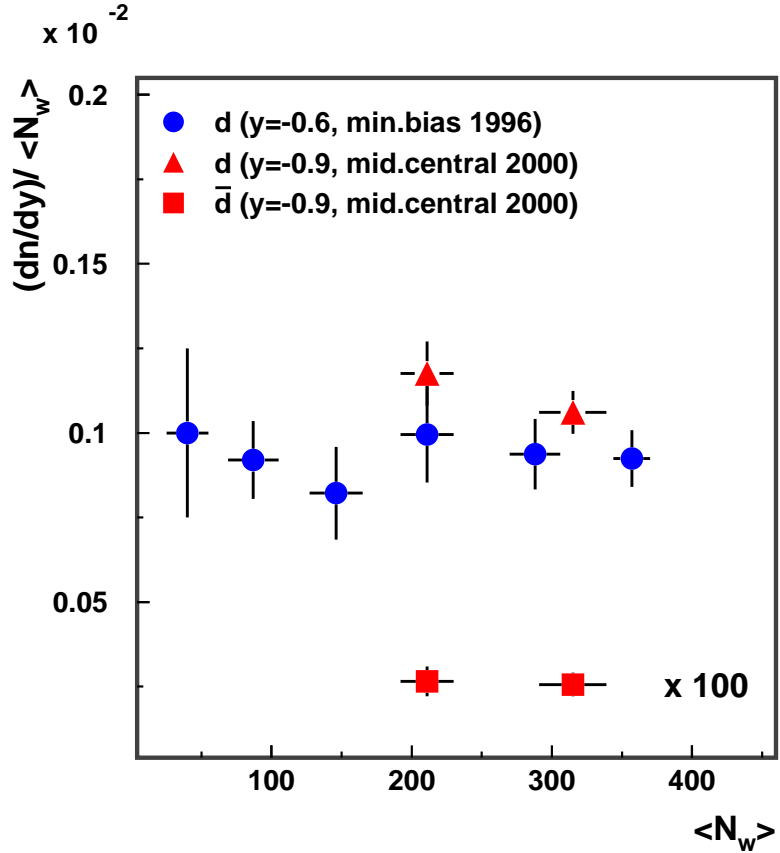}
\caption{(Color online) $dN/dy$ divided by the average number of
wounded nucleons $\langle N_w \rangle$ 
for $\overline{d}$ and $d$ for Pb+Pb collisions at 158\agev.
 }
\label{fig:dbar_yield}
\end{figure}

In Fig.~\ref{fig:dbar_yield} the invariant yield of $\overline{d}$ and $d$
normalized to the number of wounded nucleons $\langle N_w \rangle$ is plotted
as a function of $\langle N_w \rangle$.
The previously published NA49 results for deuterons from
the year 1996 minimum bias data set~\cite{clust_na49} (blue circles)
are also shown.
Within the experimental errors, the centrality dependence (if any) of the
$\overline{d}$ yield per $N_w$ appears to be weak in mid-central
Pb+Pb collisions. Also the normalized yield for $d$ shows little variation
in the entire centrality range. This behavior can be understood
as an indication of some degree of saturation in the
density distribution of both nucleons and antinucleons achieved in mid-central Pb+Pb
collisions at the top SPS energy.

\subsection{Coalescence}

A combined analysis of the invariant yield spectra of composites
and (anti)nucleons has been performed in the framework
of a coalescence approach. This approach \cite{coal1,coal2} relates
the invariant yield of $Z$-charged $A$-clusters ($A$ denotes the atomic
mass number and $P$ the momentum) to the product of the yields of protons ($p$) and
neutrons ($n$) as
\begin{equation}
\label{eq:b2}
E_A\frac{\der^3N_A}{\der^3P}=B_A\left(E_p
\frac{\der^3N_p}{\der^3p}\right)^Z\left(E_n
\frac{\der^3N_n}{\der^3p}\right)^{A-Z}
\end{equation}
The (unmeasured) yield of neutrons is usually considered to be
equal to that of protons; such an assumption seems quite reasonable
at SPS energies~\cite{na49_trit_hel}.
Thermal models of cluster production~\cite{capusta,mekyan} predict the
coalescence parameter $B_A$ to be inversely proportional to the volume
of the particle source. Such models assume the same freeze-out conditions
for matter and antimatter in chemical and thermal equilibrium.
More advanced (hydrodynamically motivated) calculations, which implement
collective expansion of the reaction zone within the density matrix
approach~\cite{coal_heinz} demonstrated a close relation of the characteristic
(coalescence) radii to those obtained from 2-particle correlation (HBT) analysis.

\begin{figure}
\includegraphics[width=0.6\linewidth]{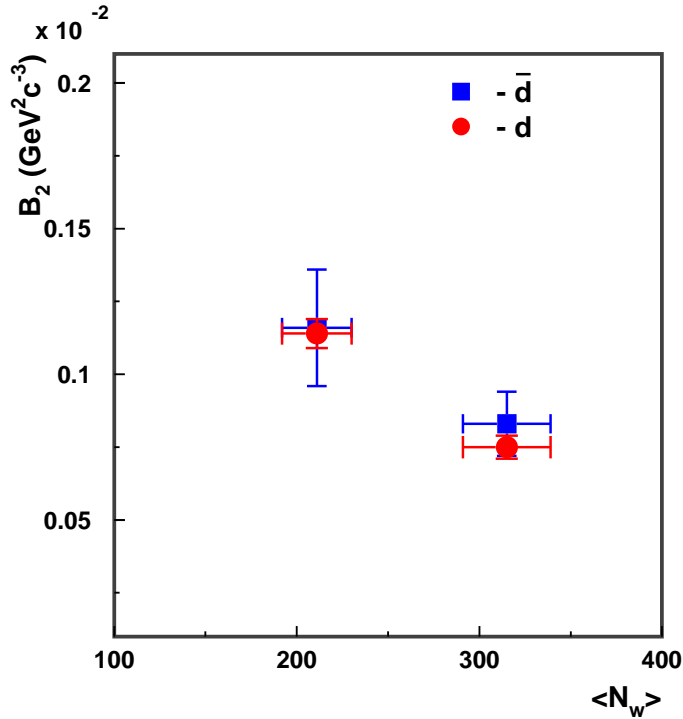}
\caption{(Color online) Coalescence parameter $B_2$ for $d$ and $\overline{d}$
calculated from Eq.~\ref{eq:b2} in centrality selected Pb+Pb collisions at 158\agev.}
\label{fig_b2_nw}
\end{figure}

%%
%%% (Vadim) added 2.08.2010
Since there is no overlap between the TOF acceptances for (anti)protons and (anti)deuterons
(the rapidity coverage of the NA49 TOF detector at 158\agev\ is $-1.2 < y < -0.6$
for $d$($\overline{d}$) and $-0.5 < y < -0.1$ for $p$($\overline{p}$)), 
the (anti)proton spectra measured by TOF were extrapolated
to $y=-0.9$. The normalization scaling factors for the $p_t$ spectra were
determined using a Gaussian rapidity distribution for antiprotons and a parabolic
function for protons. The function parameters were fitted to the experimental
data~\cite{na49_ppbar_dedx}. The shape of the $p_t$ spectra, however, remained
unchanged; this assumption is based on the results of~\cite{na49_ppbar_dedx}
which demonstrate that within $|y| < 1$ there is no significant variation
in the transverse shape parameters of $p$($\overline{p}$) production.
In Fig.~\ref{fig_b2_nw} (numerical values are given in Table~\ref{table:tab3})
the coalescence factors $B_2$ for $d$($\overline{d}$) as 
calculated from \Eq{eq:b2} and averaged over the $p_t$ interval 0~-~0.9~\gevc\
are plotted as a function of $\langle N_w \rangle$ for two centrality selected event samples.
The $B_2$ parameters for the $A$=2 nuclei and antinuclei agree within the errors; so the
most likely conclusion is that the effective coalescence volumes for clusters
and anticlusters are similar. %{\bf
 The strong centrality dependence of $B_2$ might be
explained as an increase of the tranverse size of the source in more central collisions
since in the centrality range under study $B_2$ approximately scales inversely with $N_w$.
%}.  
\begin{table}[tbh]%[H] add [H] placement to break table across pages
\caption{\label{table:tab3} Coalescence parameters $B_2$ as calculated from Eq.~\ref{eq:b2} 
and averaged over the $p_t$ interval 0~-~0.9~\gevc\ for $d$ and $\overline{d}$ 
in centrality selected Pb+Pb collisions at 158\agev\ in the 
rapidity interval $-1.2<y<-0.6$. Errors are statistical only.}
\begin{ruledtabular}
\begin{tabular}{lll}
        \multicolumn{1}{c}{centrality}
        &\multicolumn{1}{c}{0~-~12.5\%}
        &\multicolumn{1}{c}{12.5~-~23.5\%} \\
\hline
 $\langle N_w \rangle$ & $315\pm 24$ & $211\pm 19$\\
 $B_2$ (GeV$^2c^{-3}$) deuterons            & $(7.5\pm 0.4)\cdot 10^{-4}$ & $(11.4\pm 0.5)\cdot 10^{-4}$\\
 $B_2$ (GeV$^2c^{-3}$) antideuterons & $(8.3\pm 1.1)\cdot 10^{-4}$ & $(11.6\pm 2.0)\cdot 10^{-4}$\\
\end{tabular}
\end{ruledtabular}
\end{table}

\begin{figure}
\includegraphics[width=0.6\linewidth]{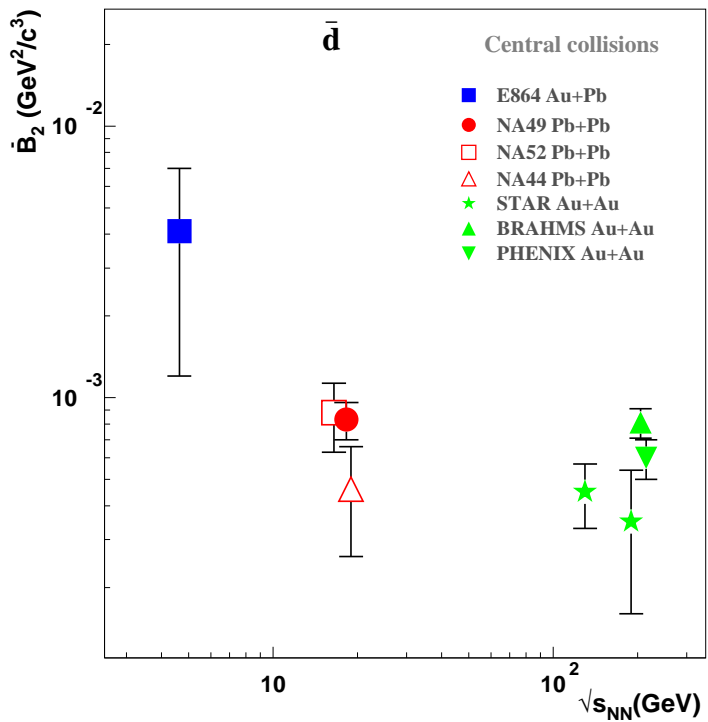}
\caption{(Color online) Coalescence parameters $B_2$ for
$\overline{d}$ in central A+A interactions at different collision energies.}
\label{fig:b2}
\end{figure}

Our measurement of $B_2$ along with other recent experimental results
on the coalescence parameter for
antideuterons~\cite{dbar_e864,dbar_na44,dbar_star,dbar_phenix,dbar_brahms}
is shown in Fig.~\ref{fig:b2}.
Taking into account the variety of experimental conditions for the $B_2$ measurements
(energy, centrality and $p_t$ interval), the experimental data indicate that
there is no substantial decrease of the coalescence parameter in the
region of \sqrts\ from 17.3 to 200~GeV. This implies that the transverse
size of the emitting source for $\overline{d}$ at kinetic freeze-out
depends only weakly on \sqrts\ in this energy domain. 

\section{Summary}

In this paper we present results on $\overline{d}$ and $d$ production in
the 23.5~\% most central Pb+Pb collisions at 158$A$~GeV.
Antideuterons were measured in the rapidity range $-1.2 < y < -0.6$ and
transverse momentum interval $0 < p_t < 0.9$~\gevc. A qualitative
comparison of the $p_t$ distributions for $\overline{d}$ and
$d$ showed close similarity of spectral shapes at all studied
collision centralities. Thus no obvious effects of annihilation
in the fireball medium were observed. The rapidity density
normalized to the number of wounded nucleons exhibits no centrality
dependence for either $d$ or $\overline{d}$.
A difference was observed in the shapes of the
rapidity distributions for $\overline{d}$ (convex) and $d$ (concave).
The $4\pi$ yields were estimated from a Gaussian and parabolic fit
for $\overline{d}$ and $d$, respectively.
The extracted coalescence parameters for $d$ and $\overline{d}$ near
midrapidity agree with each other within errors, implying a
similar freeze-out configuration for matter and antimatter.

\begin{acknowledgments}
This work was supported by
the US Department of Energy Grant DE-FG03-97ER41020/A000,
the Bundesministerium fur Bildung und Forschung, Germany (06F~137),
the Virtual Institute VI-146 of Helmholtz Gemeinschaft, Germany,
the Polish Ministry of Science and Higher Education (1~P03B~006~30, 1~P03B~127~30, 
0297/B/H03/2007/33, N~N202~078735,  N~N202~078738, N~N202~204638),
the Hungarian Scientific Research Foundation (T032648, T032293, T043514),
the Hungarian National Science Foundation, OTKA, (F034707),
the Bulgarian National Science Fund (Ph-09/05),
the Croatian Ministry of Science, Education and Sport (Project 098-0982887-2878)
and Stichting FOM, the Netherlands.
\end{acknowledgments}

%==============================================

\newpage
\section{Appendix}

Tables~\ref{table:tab4},\ref{table:tab5},\ref{table:tab6} and \ref{table:tab7} list the 
invariant transverse momentum spectra $d^2N/(p_tdp_tdy)$~in GeV$^{-2}c^2$ 
for $d$, $\overline{d}$, $p$ and $\overline{p}$, respectively, 
in centrality selected Pb+Pb collisions at 158$A$~GeV.

\begin{table}[tbh]%[H] add [H] placement to break table across pages
\caption{\label{table:tab4} Transverse momentum spectra 
$d^2N/(p_tdp_tdy)$~in GeV$^{-2}c^2$ for deuterons
in centrality selected Pb+Pb collisions at 158$A$ GeV
averaged over the rapidity interval $-1.2 < y < -0.6$.}
\begin{ruledtabular}
\begin{tabular}{l|lll}
        \multicolumn{1}{c}{  $p_t$  }
        &\multicolumn{1}{c}{0-12.5\%}
        &\multicolumn{1}{c}{12.5-23.5\%}
        &\multicolumn{1}{c}{0-23.5\%} \\
\hline
0.10 & $(3.568\pm 0.13)\cdot10^{-1}$ & $(2.780\pm 0.12)\cdot10^{-1}$ & $(3.205\pm 0.12)\cdot10^{-1}$ \\ 
0.30 & $(3.360\pm 0.97)\cdot10^{-1}$ & $(2.667\pm 0.089)\cdot10^{-1}$ & $(3.041\pm 0.093)\cdot10^{-1}$ \\
0.50 & $(3.097\pm 0.12)\cdot10^{-1}$ & $(2.520\pm 0.11)\cdot10^{-1}$ & $(2.832\pm 0.11)\cdot10^{-1}$ \\
0.70 & $(2.814\pm 0.15)\cdot10^{-1}$ & $(2.204\pm 0.14)\cdot10^{-1}$ & $(2.533\pm 0.14)\cdot10^{-1}$ \\
0.90 & $(2.301\pm 0.13)\cdot10^{-1}$ & $(1.830\pm 0.12)\cdot10^{-1}$ & $(2.084\pm 0.12)\cdot10^{-1}$ \\
1.10 & $(1.865\pm 0.11)\cdot10^{-1}$ & $(1.424\pm 0.095)\cdot10^{-1}$ & $(1.662\pm 0.10)\cdot10^{-1}$ \\
1.30 & $(1.468\pm 0.11)\cdot10^{-1}$ & $(1.076\pm 0.090)\cdot10^{-1}$ & $(1.288\pm 0.10)\cdot10^{-1}$ \\
1.50 & $(0.991\pm 0.085)\cdot10^{-1}$ & $(0.776\pm 0.081)\cdot10^{-1}$ & $(0.892\pm 0.083)\cdot10^{-1}$ \\
1.70 & $(0.732\pm 0.076)\cdot10^{-1}$ & $(0.521\pm 0.066)\cdot10^{-1}$ & $(0.634\pm 0.072)\cdot10^{-1}$ \\
1.90 & $(0.454\pm 0.057)\cdot10^{-1}$ & $(0.345\pm 0.053)\cdot10^{-1}$ & $(0.404\pm 0.055)\cdot10^{-1}$ \\
2.10 & $(0.339\pm 0.053)\cdot10^{-1}$ & $(0.264\pm 0.045)\cdot10^{-1}$ & $(0.306\pm 0.049)\cdot10^{-1}$ \\
\end{tabular}
\end{ruledtabular}
\end{table}

\begin{table}[tbh]%[H] add [H] placement to break table across pages
\caption{\label{table:tab5} Transverse momentum spectra 
$d^2N/(p_tdp_tdy)$~in GeV$^{-2}c^2$ for antideuterons 
in centrality selected Pb+Pb collisions at 158$A$ GeV
averaged over the rapidity interval $-1.2 < y < -0.6$.}
\begin{ruledtabular}
\begin{tabular}{l|lll}
        \multicolumn{1}{c}{$p_t$}
        &\multicolumn{1}{c}{0-12.5\%}
        &\multicolumn{1}{c}{12.5-23.5\%}
        &\multicolumn{1}{c}{0-23.5\%} \\
\hline
0.075 & $(9.33\pm 2.5)\cdot 10^{-4}$ & $(6.67\pm 2.4)\cdot10^{-4}$ &  $(8.00\pm 1.7)\cdot10^{-4}$ \\ 
0.225 & $(8.84\pm 1.8)\cdot 10^{-4}$ & $(7.42\pm 1.8)\cdot10^{-4}$ &  $(7.91\pm 1.3)\cdot 10^{-4}$ \\
0.375 & $(9.01\pm 2.2)\cdot 10^{-4}$ & $(4.64\pm 1.6)\cdot10^{-4}$ &  $(7.07\pm 1.4)\cdot 10^{-4}$ \\
0.675 & $(4.92\pm 1.9)\cdot 10^{-4}$ & $(4.17\pm 2.0)\cdot10^{-4}$ &  $(4.21\pm 1.4)\cdot 10^{-4}$ \\
\end{tabular}
\end{ruledtabular}
\end{table}

\begin{table}[tbh]%[H] add [H] placement to break table across pages
\caption{\label{table:tab6} Transverse momentum spectra 
$d^2N/(p_tdp_tdy)$~in GeV$^{-2}c^2$ for protons
in centrality selected Pb+Pb collisions at 158$A$ GeV
averaged over the rapidity interval $-0.5 < y < -0.1$.}
\begin{ruledtabular}
\begin{tabular}{l|lll}
        \multicolumn{1}{c}{  $p_t$  }
        &\multicolumn{1}{c}{0-12.5\%}
        &\multicolumn{1}{c}{12.5-23.5\%}
        &\multicolumn{1}{c}{0-23.5\%} \\
\hline
0.05 & $(52.84\pm 0.90)$ & $(38.64\pm 0.54)$ & $(46.01\pm 0.73)$ \\ 
0.15 & $(50.74\pm 0.52)$ & $(36.82\pm 0.31)$ & $(44.04\pm 0.42)$ \\
0.25 & $(48.85\pm 0.48)$ & $(35.44\pm 0.29)$ & $(42.39\pm 0.39)$ \\
0.35 & $(44.41\pm 0.49)$ & $(31.19\pm 0.29)$ & $(38.05\pm 0.39)$ \\
0.45 & $(40.86\pm 0.49)$ & $(29.28\pm 0.29)$ & $(35.28\pm 0.39)$ \\
0.55 & $(37.11\pm 0.44)$ & $(26.06\pm 0.26)$ & $(31.79\pm 0.36)$ \\
0.65 & $(33.69\pm 0.40)$ & $(23.01\pm 0.23)$ & $(28.55\pm 0.32)$ \\
0.75 & $(28.19\pm 0.35)$ & $(18.69\pm 0.20)$ & $(23.62\pm 0.28)$ \\
0.85 & $(22.94\pm 0.30)$ & $(15.42\pm 0.18)$ & $(19.32\pm 0.24)$ \\
0.95 & $(17.72\pm 0.26)$ & $(12.09\pm 0.15)$ & $(15.01\pm 0.21)$ \\
1.05 & $(14.43\pm 0.22)$ & $(9.288\pm 0.15)$ & $(11.95\pm 0.18)$ \\
1.15 & $(11.12\pm 0.19)$ & $(7.138\pm 0.11)$ & $(9.205\pm 0.15)$ \\
1.25 & $(8.181\pm 0.16)$ & $(5.312\pm 0.089)$ & $(6.799\pm 0.12)$ \\
1.35 & $(5.874\pm 0.13)$ & $(3.950\pm 0.074)$ & $(4.947\pm 0.10)$ \\
1.45 & $(4.459\pm 0.11)$ & $(2.899\pm 0.061)$ & $(3.708\pm 0.086)$ \\
1.55 & $(3.083\pm 0.089)$ & $(1.901\pm 0.050)$ & $(2.557\pm 0.071)$ \\
1.65 & $(2.162\pm 0.079)$ & $(1.415\pm 0.045)$ & $(1.802\pm 0.063)$ \\
1.75 & $(1.535\pm 0.070)$ & $(1.028\pm 0.040)$ & $(1.291\pm 0.056)$ \\
1.85 & $(1.096\pm 0.064)$ & $(0.723\pm 0.036)$ & $(0.916\pm 0.050)$ \\
1.95 & $(0.749\pm 0.058)$ & $(0.477\pm 0.032)$ & $(0.618\pm 0.046)$ \\
2.05 & $(0.486\pm 0.051)$ & $(0.320\pm 0.029)$ & $(0.406\pm 0.040)$ \\
\end{tabular}
\end{ruledtabular}
\end{table}

\begin{table}[tbh]%[H] add [H] placement to break table across pages
\caption{\label{table:tab7} Transverse momentum spectra 
$d^2N/(p_tdp_tdy)$~in GeV$^{-2}c^2$ for antiprotons
in centrality selected Pb+Pb collisions at 158$A$ GeV
averaged over the rapidity interval $-0.5 < y < -0.1$.}
\begin{ruledtabular}
\begin{tabular}{l|lll}
        \multicolumn{1}{c}{  $p_t$  }
        &\multicolumn{1}{c}{0-12.5\%}
        &\multicolumn{1}{c}{12.5-23.5\%}
        &\multicolumn{1}{c}{0-23.5\%} \\
\hline
0.05 & $(4.819\pm 0.24)$ & $(3.507\pm 0.15)$ & $(4.221\pm 0.20)$ \\ 
0.15 & $(4.548\pm 0.14)$ & $(3.323\pm 0.086)$ & $(3.991\pm 0.11)$ \\
0.25 & $(4.044\pm 0.12)$ & $(3.003\pm 0.074)$ & $(3.574\pm 0.097)$ \\
0.35 & $(3.694\pm 0.12)$ & $(2.654\pm 0.074)$ & $(3.223\pm 0.100)$ \\
0.45 & $(3.076\pm 0.11)$ & $(2.395\pm 0.071)$ & $(2.774\pm 0.094)$ \\
0.55 & $(2.743\pm 0.11)$ & $(2.054\pm 0.066)$ & $(2.434\pm 0.088)$ \\
0.65 & $(2.377\pm 0.099)$ & $(1.679\pm 0.059)$ & $(2.060\pm 0.079)$ \\
0.75 & $(1.871\pm 0.085)$ & $(1.377\pm 0.052)$ & $(1.649\pm 0.069)$ \\
0.85 & $(1.470\pm 0.074)$ & $(1.073\pm 0.045)$ & $(1.291\pm 0.060)$ \\
0.95 & $(1.226\pm 0.065)$ & $(0.833\pm 0.038)$ & $(1.046\pm 0.052)$ \\
1.05 & $(0.907\pm 0.056)$ & $(0.659\pm 0.033)$ & $(0.794\pm 0.045)$ \\
1.15 & $(0.743\pm 0.050)$ & $(0.500\pm 0.029)$ & $(0.632\pm 0.040)$ \\
1.25 & $(0.562\pm 0.045)$ & $(0.411\pm 0.027)$ & $(0.494\pm 0.037)$ \\
1.35 & $(0.401\pm 0.041)$ & $(0.304\pm 0.025)$ & $(0.358\pm 0.033)$ \\
1.45 & $(0.302\pm 0.038)$ & $(0.231\pm 0.024)$ & $(0.271\pm 0.031)$ \\
\end{tabular}
\end{ruledtabular}
\end{table}


\section*{References}
\begin{thebibliography}{10}
\bibitem{antip_heinz}U.~Heinz {\it et al.}, J. Phys. G {\bf 12} (1986) 1237.
\bibitem{antip_koch}P.~Koch {\it et al.}, Mod. Phys. Lett. A {\bf 3} (1988) 737.
\bibitem{antip_ellis}J.~Ellis {\it et al.}, Phys. Lett. B {\bf 233} (1989) 223.
\bibitem{becat}F.~Becattini {\it et al.}, Phys. Rev. C {\bf 73} (2006) 044905.
\bibitem{antib_bleicher} M.~Bleicher {\it et al.}, Phys. Lett. B {\bf 485} (2000) 133.
\bibitem{antip_rapp} R.~Rapp and E.~V.~Shuryak, Nucl. Phys. A {\bf 698} (2002) 587.
\bibitem{antip_gavin}S.~Gavin {\it et al.}, Phys. Lett. B {\bf 234} (1990) 175.
\bibitem{clust_e878}M.~J.~Bennett {\it et al.} (E878 Collaboration), Phys. Rev. C {\bf 58} (1998) 1155.
\bibitem{clust_na52}G.~Ambrosini {\it et al.} (NA52 Collaboration), Phys. Lett. B {\bf 417} (1998) 202.
\bibitem{clust_e802}L.~Ahle {\it et al.} (E802 Collaboration), Phys. Rev. C {\bf 60} (1999) 064901.
\bibitem{clust_e864}T.~A.~Armstrong {\it et al.} (E864 Collaboration), Phys. Rev. C {\bf 61} (2000) 064908.
\bibitem{clust_e877}J.~Barrette {\it et al.} (E877 Collaboration), Phys. Rev. C {\bf 61} (2000) 044906.
\bibitem{clust_na49} T.~Anticic {\it et al.} (NA49 Collaboration), Phys. Rev. C {\bf 69} (2004) 024902.
\bibitem{hyper_ags}T.~A.~Armstrong {\it et al.} (E864 Collaboration), Phys. Rev. C {\bf 70} (2004) 024902.
\bibitem{hyper_star}STAR Collaboration, Science {\bf 328} (2010) 58.
\bibitem{coal1}S.~T.~Butler and C.~A.~Pearson, Phys. Rev. {\bf 129} (1963) 836.
\bibitem{coal2}A.~Schwarzschild and C.~Zupancic,  Phys. Rev. {\bf 129} (1963) 854.
\bibitem{na49_hbt} C.~Alt {\it et al.} (NA49 Collaboration), Phys. Rev. C {\bf 77} (2008) 064908.
\bibitem{slope_polleri}A.~Polleri {\it et al.}, Phys. Lett. B {\bf 419} (1998) 19.
\bibitem{antib_bleicher_2} M.~Bleicher {\it et al.}, Phys. Lett. B {\bf 361} (1995) 10.
\bibitem{mrozhin}S.~Mrowczynski, Phys. Lett. B {\bf 277} (1992) 43.
\bibitem{coal_heinz} R.~Scheibl and U.~Heinz, Phys. Rev. C {\bf 59} (1999) 1585.
%\bibitem{e864_1}T.~A.~Armstrong {\it et al.}, Phys. Rev. C {\bf 61} (2000) 064908.
\bibitem{na49_ppbar} C.~Alt {\it et al.} (NA49 Collaboration), Phys. Rev. C {\bf 73} (2006) 044910.
%\bibitem{anti1} S.~Gavin {\it et al.}, Phys. Lett. B {\bf 234} (1990) 175.
%\bibitem{anti2} J.~Schaffner {\it et al.}, Z. Phys. A {\bf 341} (1991) 47.
%\bibitem{anti3} H.~Sorge {\it et al.}, Phys. Lett. B {\bf 289} (1992) 6.
%\bibitem{na49_pd} T.~Anticic {\it et al.} (NA49 Collaboration), Phys. Rev. C {\bf 69} (2004) 024902.
\bibitem{na49_setup} S.~V.~Afanasiev {\it et al} (NA49 Collaboration), Nucl. Instrum. Meth. A{\bf 430} (1999) 210.
%\bibitem{Wroblew} A.~Wroblewski, Acta Phys. Pol. B {\bf 16} (1985) 379.
\bibitem{venus}K.~Werner, Z.~Physik C {\bf 42} (1989) 85.
\bibitem{na49_centrality} Details can be found in: https://edms.cern.ch/file/885329/1/vetocal2.pdf.
\bibitem{na49_lam_energy} C.~Alt {\it et al.} (NA49 Collaboration), Phys. Rev. C {\bf 78} (2008) 034918.
\bibitem{na49_lam_size} T.~Anticic {\it et al.} (NA49 Collaboration), Phys. Rev. C {\bf 80} (2009) 034906.
\bibitem{urqmd}M.~Bleicher {\it et al.}, J. Phys. G {\bf 25} (1999) 1859.
\bibitem{anti_loss}F.~Wang, J. Phys. G {\bf 27} (2001) 283.
\bibitem{na49_trit_hel}V.~I.~Kolesnikov (for the NA49 Collaboration),
J. Phys. Conf. Ser. {\bf 110} (2008) 032010.
\bibitem{na49_ppbar_dedx} T.~Anticic {\it et al.} (NA49 collaboration), Phys. Rev. C {\bf 83} (2011) 014901.
\bibitem{becatini_private}F.~Becattini, private communication.
\bibitem{capusta}J.~Kapusta, Phys. Rev. C {\bf 21} (1980) 1301.
\bibitem{mekyan}A.~Z.~Mekjian, Phys. Rev. C {\bf 17} (1978) 1051.
\bibitem{dbar_e864} T.~A.~Armstrong {\it et al.} (E864 Collaboration), Phys. Rev. Lett. {\bf 85} (2000) 2685.
\bibitem{dbar_na44} I.~G.~Bearden {\it et al.} (NA44 Collaboration), Phys. Rev. Lett. {\bf 85} (2000) 2681.
\bibitem{dbar_star} C.~Adler {\it et al.} (STAR Collaboration), Phys. Rev. Lett. {\bf 87} (2001) 262301.
\bibitem{dbar_phenix} S.S.~Adler {\it et al.} (PHENIX Collaboration), Phys. Rev. Lett. {\bf 94} (2005) 122302.
\bibitem{dbar_brahms}I.~Arsene {\it et al.} (BRAHMS Collaboration), preprint [nucl-ex] 1005.5427 (2010).
\end{thebibliography}
\end{document}